\documentclass[twocolumn,trackchanges]{aastex701}

\usepackage{booktabs} 
\usepackage{amsmath} 
\newenvironment{DIFnomarkup}{}{}

\begin{document}

\title{A tale of three tails: A misaligned streamer and mysterious structures around [BHB2007]1}

\author[orcid=0000-0002-9959-1933]{Aashish Gupta}
\affiliation{Department of Astronomy, University of Virginia, Charlottesville, VA 22904, USA}
\email[show]{aashishgupta@virginia.edu}  

\author[orcid=0000-0001-5073-2849]{Antonio S. Hales}
\affiliation{National Radio Astronomy Observatory, 520 Edgemont Road, Charlottesville, VA 22903-2475, United States of America}
\affiliation{Joint ALMA Observatory, Avenida Alonso de C\'ordova 3107, Vitacura 7630355, Santiago, Chile}
\affiliation{Millennium Nucleus on Young Exoplanets and their Moons (YEMS), Chile}
\email{ahales@nrao.edu}

\author[0000-0003-2076-8001]{L. Ilsedore Cleeves}
\affiliation{Department of Astronomy, University of Virginia, Charlottesville, VA 22904, USA}
\email{lic3f@virginia.edu}

\author[0000-0002-7945-064X]{Felipe Alves}
\affiliation{Institut de Radioastronomie Millimétrique, 300
rue de la Piscine, Domaine Universitaire, 38406 Saint-Martin d'Hères, France}
\email{alves@iram.fr}

\author[]{Trisha Bhowmik}
\affiliation{Institute of Astrophysical Studies, Faculty of Engineering and Sciences, Diego Portales University, Av. Ejército 441, Santiago, Chile}
\affiliation{Millennium Nucleus on Young Exoplanets and their Moons (YEMS), Chile}
\email{trisha.bhowmik@mail.udp.cl}

\author{Nicolás Cuello}
\affiliation{Univ. Grenoble Alpes, CNRS, IPAG, 38000 Grenoble, France}
\email{nicolas.cuello@univ-grenoble-alpes.fr}

\author{Josep M. Girart}
\affiliation{Institut de Ciències de l’Espai (ICE), CSIC, Can Magrans s/n, Cerdanyola del Vallès, E-08193 Catalonia, Spain}
\affiliation{Institut d’Estudis Espacials de Catalunya (IEEC), E-08034 Barcelona, Catalonia, Spain}
\email{girart.astro@gmail.com}

\author{Zhi-Yun Li}
\affiliation{Department of Astronomy, University of Virginia, Charlottesville, VA 22904, USA}
\email{zl4h@virginia.edu}

\author{Anna Miotello}
\affiliation{European Southern Observatory, Karl-Schwarzschild-Str. 2, 85748 Garching bei München, Germany}
\email{amiotell@eso.org}

\author[]{Zhaohuan Zhu}
\affiliation{Department of Physics and Astronomy, University of Nevada, 4505 South Maryland Parkway, Las Vegas, NV 89154, USA}
\email{zhaohuan.zhu@unlv.edu}

\author[orcid=0000-0002-5903-8316]{Alice Zurlo}
\affiliation{Instituto de Estudios Astrof\'isicos, Facultad de Ingenier\'ia y Ciencias, Universidad Diego Portales, Av. Ej\'ercito Libertador 441, Santiago, Chile}
\affiliation{Millennium Nucleus on Young Exoplanets and their Moons (YEMS), Chile}
\email{alice.zurlo@mail.udp.cl}


\begin{abstract}

Recent discoveries of streamer-like structures around protostellar sources challenge the traditional picture of isolated, axisymmetric star formation. Here, we present new ALMA observations of [BHB2007]1, a flat-spectrum source connected to at least three such elongated structures. 
Two of these features are symmetrically located to the north and south of the disk, with velocities aligned with the disk on their respective sides.
However, their unbound kinematics and curved morphology make it difficult to determine their origin. 
Possible explanations include outflows, interactions with the nearby BHB2 system, and hyperbolic infall, but none fully account for all observed properties.
In contrast, a newly identified collimated structure to the west shows clear evidence of gravitationally bound infall. 
Estimates of its mass, mass infall rate, and angular momentum suggest that this infalling streamer would roughly double the mass budget available to form planets and tilt the disk by a few tens of degrees. 
Furthermore, its misalignment with the midplane of the disk and the lack of diffuse envelope emission indicate that the streamer may have formed due to gravitational capture of cloud material unrelated to the source’s natal core. Together, these findings support a more dynamic picture of star formation, one where environmental interactions continue to shape conditions for building planetary systems.


\end{abstract}



\section{Introduction}
\label{sec:intro}

The textbook description of star formation describes stars forming through the gravitational collapse of dense gaseous cores, assumed to be isolated and axisymmetric \citep[e.g.,][]{Shu1977,Terebey1984}. While these young stellar objects (YSOs) are still embedded within their natal gaseous envelopes, a disk of material forms around them. Within the disks around these Class I YSOs, planet formation is expected to begin \citep[e.g.,][]{Tychoniec2020}. As the protostars gain most of their final mass and the surrounding gas envelopes disperse, they transition into Class II systems, where disks are expected to evolve in isolation to form planetary systems.

Numerous recent observational and theoretical studies are challenging this traditional view of axisymmetric Class 0 and I sources, and isolated Class II disks \citep{Pineda2023}. 
In particular, sub-mm and scattered light observations are revealing asymmetric channels of infalling material, often referred to as streamers, around very young Class 0 \citep[e.g.,][]{Tobin2012,Pineda2020,Thieme2022,Hoff2023,Sharma2025}, intermediate Class I \citep[e.g.,][]{Yen2014,Valdivia-Mena2024,Hsieh2023,Mercimek2023,Cacciapuoti2023,Flores2023,Hales2024,Tanious2024} and even evolved Class II \citep[e.g.,][]{Tang2012,Akiyama2019,Huang2020,Huang2021,Huang2022,Huang2023,Gupta2023,Garufi2022,Garufi2024, Ginski2024} sources. 
They are expected to be direct counterparts to the infalling streamers commonly observed in global simulations of star-forming regions \citep[e.g.,][]{Padoan2014,Haugbolle2018,Kuznetsova2019,Lebreuilly2021,Pelkonen2021,Kuffmeier2023}. 

[BHB2007]-1 or 2MASS J17110392-2722551, hereafter BHB1, is one such YSO expected to be interacting with its environment \citep{Alves2020}. \cite{Alves2020} reported Atacama Large Millimeter/submillimeter Array (ALMA) $^{12}$CO(2--1) observations of this YSO, which show two prominent elongated structures that appear to be connected to it.
Based on modeling its Spectral Energy Distribution (SED), BHB1 has been classified as a ``flat-spectrum" YSO \citep{Forbrich2009}. Although this observational class has no direct counterpart in the traditional theoretical stages of star formation, it is generally expected to be a source transitioning from Class I to Class II stage. We note that the classification of individual YSOs is quite uncertain, and an evolved source can appear to be younger due to effects of disk inclination \citep[e.g.,][]{Dunham2014} and late-stage infall of material \citep[e.g.,][]{Kuffmeier2023}. 
BHB1 is located in Barnard 59 (B59), the only site of active star formation in the quiescent Pipe nebula \citep{Brooke2007}.
Gaia DR3 parallax suggests that the source is $\sim163$~pc away \citep[][]{Dzib2018}.

Besides the elongated structures, \cite{Alves2020} further reported a symmetric gap in the 1.3 mm continuum observations and within this gap, warm CO and centimetre-wavelength continuum excess. These features were interpreted as being caused by a 4--70~M$_{\rm Jup}$ mass companion. This was further supported by near-infrared (NIR) observations reported in \cite{Zurlo2021}, where the mass of the expected companion was further constrained to 37--47~M$_{\rm Jup}$. 

Here we present new ALMA observations of this source, which provide further information on the extent and dynamics of the large-scale structures around BHB1. These observations are described in Section \ref{sec:obs}. We derive properties of these structures in Section \ref{sec:analysis} and discuss our results in Section \ref{sec:discussion}. Finally, Section \ref{sec:sum} summarizes our key findings. 

\section{Observations} \label{sec:obs}

The observations of BHB1  presented here were carried out by the ALMA 12-m (C-1 configuration) and 7-m (also known as Atacama Compact Array; ACA) arrays to obtain spatial coverage from $\sim1\arcsec$ to $\sim30\arcsec$ (project code 2023.1.01065.S, PI: I. Cleeves). The
log of the observations and
array characteristics is reported in Table~\ref{tab:log1}. The correlators were configured to
observe CO(2--1) (230.538~GHz), $^{13}$CO(2--1) (220.399~GHz), and C$^{18}$O(2--1) (219.560~GHz) emission lines in a single Band 6 spectral setting (line frequencies from Splatalogue\footnote[1]{https://splatalogue.online}). They provided spectral resolution of $\sim141$~kHz, which corresponds to a velocity resolution of $\sim0.17$~km~s$^{-1}$.

\begin{DIFnomarkup}
\begin{deluxetable*}{lcccccccccccc}
\tablecaption{Summary of ALMA Observations \label{tab:log1}}
\tablewidth{700pt}
\tabletypesize{\scriptsize}
\tablehead{
\colhead{Array} & 
\colhead{\# Executions} &
\colhead{Mean PWV} &
\colhead{Baseline} &
\colhead{Avg. elevation} &
\colhead{Time on Source} &
\colhead{Line sensitivity (10 km s$^{-1}$)} &
\colhead{AR} &
\colhead{MRS} &
\\
\colhead{} & 
\colhead{} & 
\colhead{ [mm]} &
\colhead{Max-Min [m]} &
\colhead{[deg]} &
\colhead{[min]}&
\colhead{[mJy~beam$^{-1}$]}&
\colhead{[\arcsec]} & 
\colhead{[\arcsec]} 
}
\startdata
12m & 2 & 2.7 &  313.7--15.1 & 77.6 & 50 & 1.2 &1.1 & 11.3 \\
7m & 4  & 1.8 &  48.9--8.9      & 79.3 & 267 & 6.8 &5.4 & 29.0\\
\hline             
\enddata
\tablecomments{12m (C1) observations were taken on 2024-03-17 and 2024-04-13. 7m (ACA) observations were taken on 2024-06-14, 2024-06-27, and 2024-06-28. PWV refers to the precipitable water vapor during the observations. AR and MRS denote the angular resolution and maximum recoverable scale, respectively.}
\end{deluxetable*}
\end{DIFnomarkup}

All data were calibrated with the ALMA Science
Pipeline (pipeline version 2023.1.0.124) in CASA 6.5.4.9 \citep{CASA2022}. 
The AutoSelfCal Version 1.0 module, included in CASA 6.5.4-9-pipeline-2023.1.0.124, was employed for self-calibration. The visibilities were successfully self-calibrated by the pipeline, resulting in an increase in the signal-to-noise ratio (S/N) by factors of approximately 3 and 6 for the 12-m and 7-m Arrays, respectively.


For the analysis presented here, we primarily rely on imaging done by combining 12-m and 7-m (C1+ACA) observations with the {\sc tCLEAN} task. Unless stated otherwise, these are the default images being referred to within the manuscript. We also image 7-m (standalone ACA) observations separately to get emission over the larger field-of-view, and these images are used almost exclusively for the mass estimation discussed in Section \ref{sec:mass}. 
For both sets of images, continuum subtraction in the visibility domain was performed before imaging using {\sc UVCONTSUB} and ``auto-multithresh" masking with default parameters (noisethreshold=5, sidelobethreshold=3, lownoisethreshold=1.5) used for imaging with {\sc tCLEAN}. 
For both images, we used briggs weighting scheme, with robust value of -2 (close to uniform weighting) for C1+ACA images and 0.5 for standalone ACA images, to find a balance between sensitivity and angular resolution. 
For the C1+ACA images, we used `mosaic' gridder instead of the default `standard' gridder, as it accounts for the different primary beam sizes of the 12 m and 7 m antennas.
Properties of all the images presented in this manuscript are summarized in Table \ref{tab:log2}.

\begin{figure*}[htbp]
  \centering
  \includegraphics[width=0.99\textwidth]{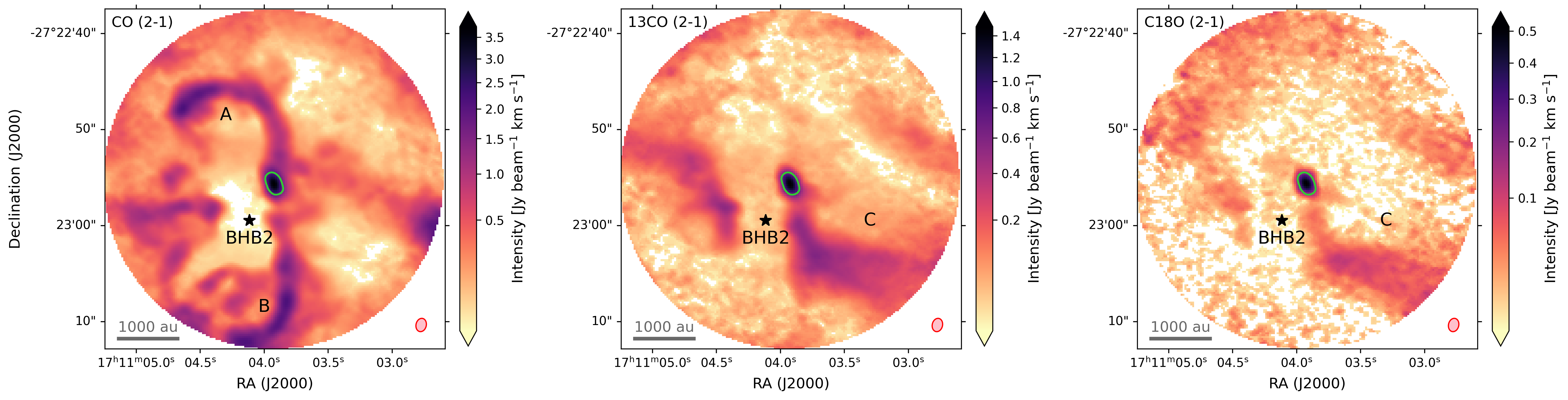}
    \caption{Integrated intensity (moment 0) maps of all three CO isotopologue emission lines imaged for this analysis: CO (2--1) (left), $^{13}$CO (2--1) (center), and C$^{18}$O (2--1) (left).
    The horizontal grey lines in the bottom-left corners represent the length scales of 1000 au, and the pink ellipses in the bottom-right corners represent the beam size.
    Green contour denote the continuum emission (5~$\sigma$) from the protoplanetary disks.
    Location of structures A, B, and C, and the nearby source BHB2 are marked on the maps. 
    }
  \label{fig:m0}
\end{figure*}

We use primary beam-corrected images for all analysis presented here, as they provide more accurate flux measurements across the field. However, since the primary beam response decreases with distance from the pointing center, the correction amplifies the noise toward the edges, making it difficult to use corrected images for analysis of structures spanning the whole field-of-view. To mitigate this, we generate a noise masks (e.g., 2$\sigma$ for Fig. \ref{fig:m0}, 3$\sigma$ for Fig. \ref{fig:aca_m0}) from the corresponding uncorrected images, where the noise is spatially uniform, and apply it to the corrected ones. This is effectively having a noise threshold that is radially increasing and azimuthally constant, to account for the radial drop in sensitivity. 



\begin{DIFnomarkup} 
\begin{deluxetable}{cccc}
\tablecaption{Image properties \label{tab:log2}}
\tablehead{
\colhead{Emission} & 
\colhead{Array} & 
\colhead{R.M.S.} &
\colhead{Beam}
}
\startdata
CO (2--1) & C1+ACA & 11.8~mJy~beam$^{-1}$$^*$ & $1.3\arcsec\times1.0\arcsec$ (PA = 74$^{\circ}$) \\
$^{13}$CO (2--1) & C1+ACA & 7.9~mJy~beam$^{-1}$$^*$ & $1.3\arcsec\times1.0\arcsec$ (PA = 76$^{\circ}$) \\
C$^{18}$O (2--1) & C1+ACA & 5.9~mJy~beam$^{-1}$$^*$ & $1.4\arcsec\times1.1\arcsec$ (PA = 77$^{\circ}$) \\
C$^{18}$O (2--1) & ACA & 21~mJy~beam$^{-1}$$^*$ & $7.7\arcsec\times4.4\arcsec$ (PA = -88$^{\circ}$) \\
Continuum (1.3~mm) & C1 & 0.03~mJy~beam$^{-1}$ & $1.3\arcsec\times1.0\arcsec$ (PA = -71$^{\circ}$) \\
\hline             
\enddata
\tablecomments{$^*$ over the typical channel width of 0.17~km~s$^{-1}$.}
\end{deluxetable}
\end{DIFnomarkup}

\begin{figure*}[htbp]
  \centering
  \includegraphics[width=0.99\textwidth]{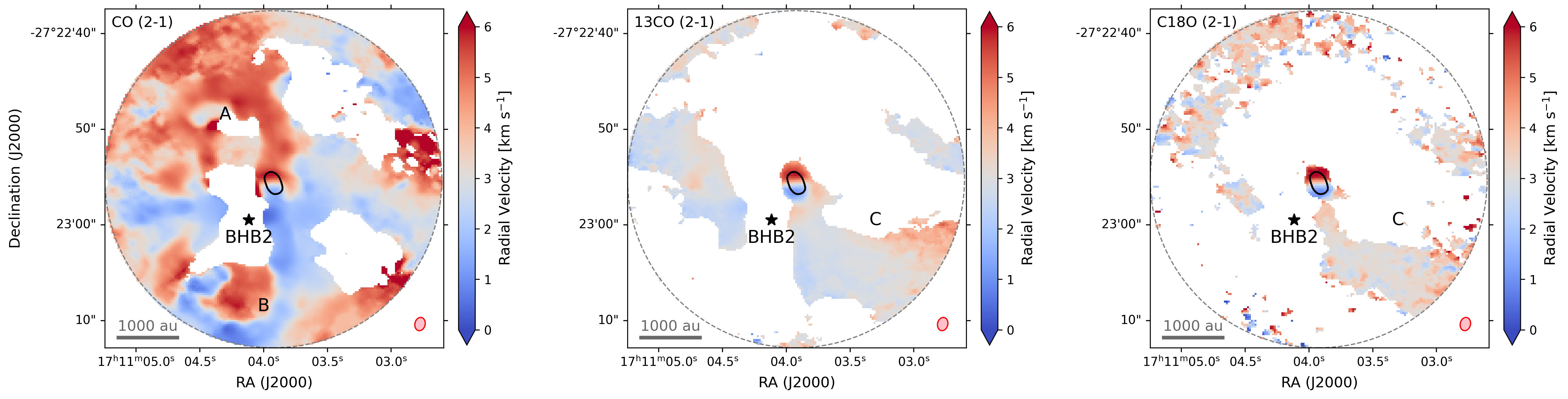}
    \caption{Intensity weighted velocity (moment 1) maps of all three CO isotopologue emission lines imaged for this analysis: CO (2--1) (left), $^{13}$CO (2--1) (center), and C$^{18}$O (2--1) (left).
    The horizontal grey lines in the bottom-left corners represent the length scales of 1000 au, and the pink ellipses in the bottom-right corners represent the beam size.
    Black contour denote the continuum emission (5~$\sigma$) from the protoplanetary disks.
    Location of structures A, B, and C, and the nearby source BHB2 are marked on the maps. 
    }
  \label{fig:m1}
\end{figure*}

\begin{figure*}[htbp]
  \centering
  \includegraphics[width=0.99\textwidth]{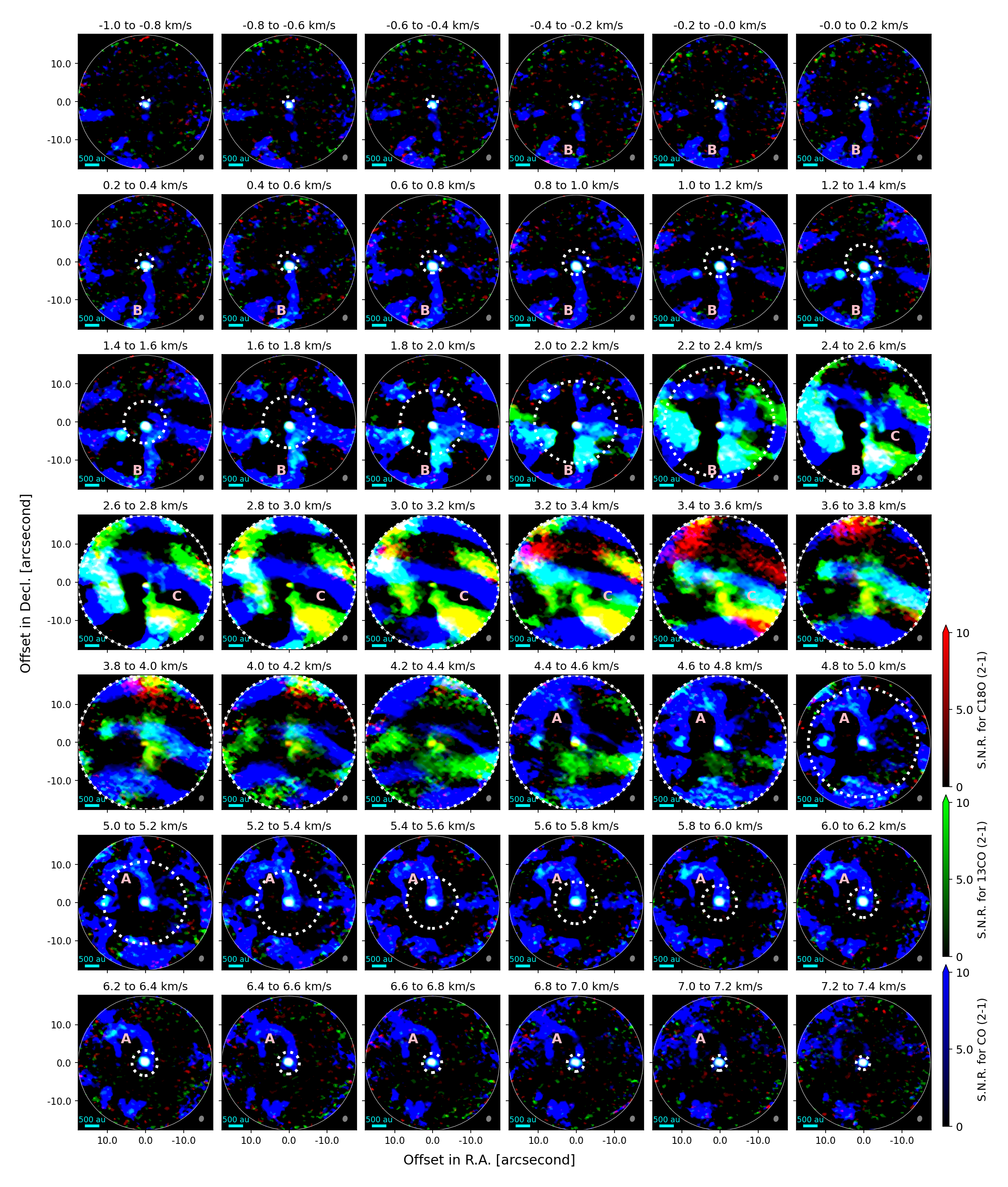}
    \caption{CO (2--1) (blue), $^{13}$CO (2--1) (green), and C$^{18}$O (2--1) (red) channel maps. 
    The colorbars indicate signal to noise ratio of the observed emission.
    Dotted white circles in channel maps denote the maximum bound radius ($r_{\text{POS,bound}}$), as defined in Section \ref{sec:analysis}.  
    Overlap of emission from different lines is seen as cyan (CO \& $^{13}$CO), yellow ($^{13}$CO \& C$^{18}$O), and white (all three lines) colors. 
    The horizontal lines in the bottom-left corners, the ellipses in the bottom-right corners, and the grey solid circles represent length scales of 500 au, beam sizes, and the field of view of the data, respectively.
    Location of structures A, B, and C are marked on the maps.
    }
  \label{fig:channels}
\end{figure*}

\section{Results and analysis} \label{sec:analysis}

The integrated emission of the three lines, shown in Figure \ref{fig:m0}, shows three prominent elongated structures connected to BHB1: two known north-south structures visible in CO (2--1) (labelled A and B) and one structure to the west visible in $^{13}$CO (2--1) and C$^{18}$O (2--1) (labelled C). The structures A and B have been previously reported in \citet{Alves2020} and were speculated to be infalling streamers. 
The structure C was not visible in previous observations because CO (2--1) is partly absorbed by foreground cloud and those observations were not sensitive enough to detect fainter $^{13}$CO and C$^{18}$O emission from this structure. 

To illustrate the dynamic nature of these structures, we show moment 1 maps in Figure \ref{fig:m1} and channel maps in Figure \ref{fig:channels}. These maps show that the structure C is emitting at velocities $\sim$2.5--3.7~km~s$^{-1}$, i.e., close to the systemic velocity of the disk ($\sim3.6$~km~s$^{-1}$). On the other hand, structure A is emitting at velocities $\gtrsim4.5$~km~s$^{-1}$ and structure B is emitting at velocities $\lesssim2.5$~km~s$^{-1}$. 

With these molecular-line observations, we obtain the relative line-of-sight (LOS) velocities ($v_{\rm LOS}$), which serve as a lower limit on the total relative velocity between the protostellar source and the surrounding gas. Likewise, we can estimate the plane-of-sky (POS) distances ($r_{\rm POS}$) between the source and the gas, providing a lower limit on the true 3D separation. Using these quantities, we can place lower limits on the specific (per unit mass) kinetic energy ($KE$) and upper limits on absolute gravitational potential energy ($PE$) of the gas relative to the protostar:
\begin{equation} \label{eq:ke_min}
    KE_{\rm min} = 0.5v_{\rm LOS}^{2}
\end{equation}
\begin{equation} \label{eq:pe_max}
    PE_{\rm max} = \frac{GM_{*}}{r_{\rm POS}}
\end{equation}
where $G$ is the gravitational constant and $M_{*}$ is the stellar mass \citep[$2.23$~M$_{\odot}$,][]{Alves2020}. Equating Equations~\ref{eq:ke_min} and \ref{eq:pe_max}, we obtain
\begin{equation}
r_{\rm POS,\text{bound}} = \frac{2GM_{*}}{v_{\rm LOS}^{2}}
\end{equation}
Here, $r_{\rm POS,\text{bound}}$ represents the projected distance beyond which $KE_{\rm min} > PE_{\rm max}$. In other words, material located beyond this radial distance is not bound to the system.

These distances ($r_{\rm POS,\text{bound}}$), as a function of $v_{\rm LOS}$, are denoted as white circles in Figure \ref{fig:channels}. As most of the emission associated with structures A and B lies beyond these distances, they are expected to be unbound from the YSO and therefore, not be typical bound streamers. 
Possible explanations for their origin is discussed in Section \ref{sec:mys}. 
Structure C, on the other hand, lies completely within $r_{\rm POS,\text{bound}}$ and is thus potentially bound to the YSO and infalling onto it. A more detailed analysis of its dynamical state is required to confirm its infalling nature, as discussed in Section \ref{sec:modeling}.




\begin{deluxetable}{l c l c}
\tablecaption{Input parameters for TIPSY modeling \label{tab:input_params}}
\tablehead{
\colhead{Parameter} & \colhead{Unit} & \colhead{Description} & \colhead{Value}
}
\startdata
$M_*$ & M$_{\odot}$ & Protostellar mass \citep[from][]{Alves2020} & 2.23 \\
D & pc & Distance to the system \citep[from][]{Dzib2018}& 163 \\
$v_{\rm sys}$ & km\,s$^{-1}$ & Systemic velocity \citep[from][]{Alves2020}& 3.6 \\
vxy0\_lim & km\,s$^{-1}$ & Range of initial speeds on POS & 0 to 1.2 \\
vxy0\_step & km\,s$^{-1}$ & Resolution of initial speeds on POS & 0.03 \\
vxy\_ang0\_span & $^\circ$ & Total span for initial angles between POS velocity and R.A. axis & 90 \\
vxy\_ang0\_step & $^\circ$ & Resolution of initial angles between POS velocity and R.A. axis & 3 \\
z0\_lim & au & Range of initial distances along LOS & $-1000$ to 3000 \\
z0\_step & au & Resolution of initial distances along LOS & 40 \\
N\_elements &  & Number of bins used to discretize the streamer point cloud & 20 \\
theta\_weight &  & Relative weight of POS angles in computing distance metric along the streamer & 0
\enddata
\tablecomments{The notebook used for this analysis is public: \url{https://github.com/AashishGpta/BHB1_analysis}.}
\end{deluxetable}

\begin{figure*}[htbp]
  \centering
  \includegraphics[width=0.99\textwidth]{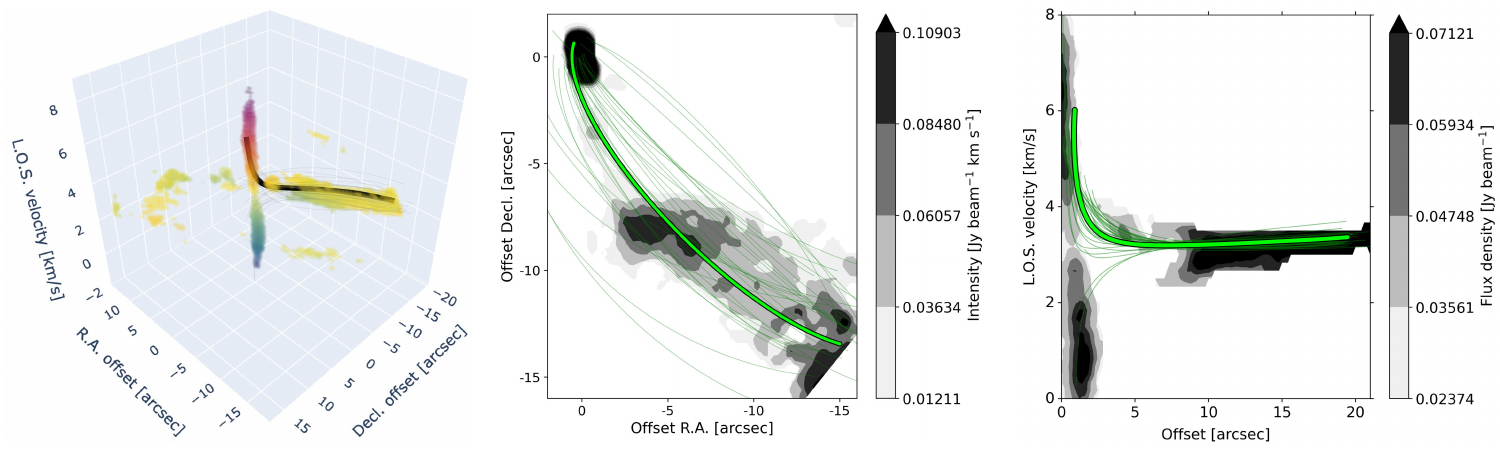}
    \caption{The best fitted trajectory from TIPSY, as discussed in the Section \ref{sec:streamer}, compared to C$^{18}$O (2--1) data.
    Left panel shows the full 3D PPV diagram of the molecular-line data, with the black line denoting the best fit.
    Center and right panels show integrated intensity (moment 0) and position-velocity diagrams for the streamer, respectively, with green curves representing the best fit. 
    Thinner curves in all panels represent trajectories generated from 30 parameter combinations, randomly sampled from a Gaussian distribution based on the uncertainties of the fitted trajectory.
    An interactive version of the left panel is available online, allowing rotation and zooming of the 3D structures. 
    }
  \label{fig:tipsy}
\end{figure*}

\subsection{Streamer dynamics} \label{sec:modeling}

To ascertain the infalling nature of structure C, which is likely bound to the system (see Figure \ref{fig:channels}), we explored if both the curved morphology (center panel, Figure \ref{fig:tipsy}) and velocity gradient (right panel, Figure \ref{fig:tipsy}) can be simultaneously explained by infalling trajectories. For this, we used a streamer fitting code, Trajectory of Infalling Particles in Streamers around Young stars\footnote[2]{https://github.com/AashishGpta/TIPSY} \citep[TIPSY,][]{Gupta2024}. TIPSY fits the observed streamer emission in the three-dimensional position-position-velocity (PPV) space, where the spatial dimensions are along R.A. and Decl., and the velocity dimension is along LOS (left panel, Figure \ref{fig:tipsy}). 
TIPSY first generates a curve-like representation of the observed streamer structure in this PPV space. Then this is compared to a generalized version of analytical trajectories for infalling material, as derived in \citet{Mendoza2009} \citep[for more details on the model, see Section 2.1,][]{Gupta2024}. The key physical assumption here is that the dynamics of the infalling material are primarily governed by the protostar’s gravity. 

For a general configuration of infalling material, we need to define the initial position ($\overrightarrow{r_{0}}$) and velocity ($\overrightarrow{v_{0}}$) of the infalling gas, relative to the protostar. The position in the POS (R.A. and Dec.) can be directly inferred from observations, while the distance along LOS remains a free parameter in the trajectory fitting (y-axes in Figure~\ref{fig:errors}). 
For velocities, observations directly constrain the LOS component, but the R.A. and Dec. components must be inferred through modeling. 
To reduce computations, the free parameters for POS velocities are parametrised as total speed on POS (x-axes in Figure \ref{fig:errors}) and initial direction on POS, which is generally well constrained by the morphology of the streamer. 
All the input parameters for TIPSY modeling, including those used to create the free parameter space, are tabulated in Table \ref{tab:input_params}.

Before the actual fitting of the streamer emission, TIPSY requires isolating the streamer emission from the rest of the cube. 
To remove the disk emission, we applied an anti-Keplerian mask following the approach of \cite{Speedie2025}. The mask was generated using the public code keplerian\_mask\footnote[3]{https://github.com/richteague/keplerian\_mask}, with parameters set to: inclination $= 76^\circ$, position angle $= 14.4^\circ$, stellar mass $= 2.23~M_\odot$, distance $= 163$~pc, systemic velocity $= 3.6$~km~s$^{-1}$, maximum radius $= 2$~arcsec, and beam convolution factor ($n_{\rm beams}$) $= 2$ (see \cite{Alves2020} and Section~\ref{sec:disk} for more details on disk parameters).
Furthermore, we restricted ourselves to emission greater than $5\sigma$ and within LOS velocities between $2$ and $5$~km~s$^{-1}$, R.A. offsets from $-20\arcsec$ to $0\arcsec$, and Dec. offsets from $-20\arcsec$ to $0\arcsec$.

The overall fitting results and uncertainties are presented in Table \ref{tab:properties} (also see Figure \ref{fig:errors}). 
For the best-fit combination of parameters, the theoretical trajectory reproduces $98\%$ of the simplified streamer data points, with a $\chi^2$ value of 21 \citep[for more details on these metrics, see Section 2.2,][]{Gupta2024}. TIPSY also enables the estimation of uncertainties on the fitted parameters, as described in Appendix~\ref{sec:erros}. The best-fit trajectories and their uncertainties are compared with observations in Figure~\ref{fig:tipsy}.
Both quantitative metrics and qualitative comparisons indicate that the best-fit model largely reproduces the observed morphology and velocity structure of the streamer, suggesting that this structure is indeed infalling on the protostellar system. However, some small deviations remain, particularly in the line-of-sight velocities close to the disk. These discrepancies may reflect additional forces, such as magnetic fields or pressure gradients, that become important at smaller scales. As these differences are mostly accounted for in our uncertainties, the TIPSY results still allow us to derive meaningful estimates of streamer properties. 


As we know the trajectory of infalling materials, we can use it to directly derive dynamical properties like specific angular momentum ($\overrightarrow{r_{0}}\times\overrightarrow{v_{0}}$) and specific total energy ($0.5v_{0}^2-GM_{*}/r_0$), which were estimated to be $725\pm438$~au~km~s$^{-1}$ and $-0.50\pm0.08$~km$^{2}$~s$^{-2}$, respectively. 
Furthermore, TIPSY provides the time taken for infalling gas to reach the point along its trajectory that is closest to the protostar ($T_{\mathrm{inf}}$), which is $17242\pm2003$~yr from the distance of $\sim3400$~au.
These dynamical properties can be combined with the mass estimate of the streamer, as discussed in Section \ref{sec:mass}, to better understand its impact on the planet-forming disk. 

\begin{figure}[htbp]
  \centering
  \includegraphics[width=0.47\textwidth]{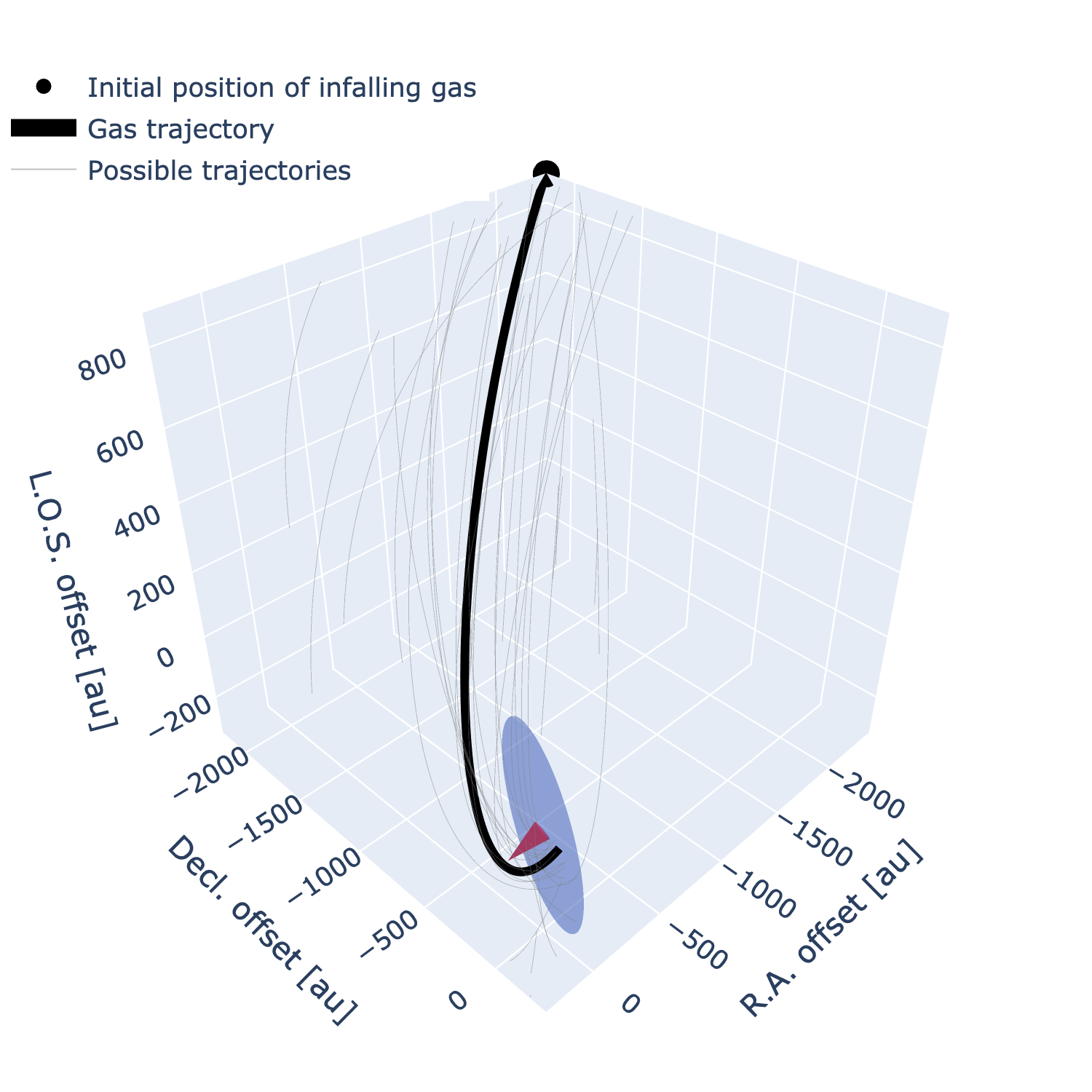}
    \caption{Isometric projection of the best-fit infalling trajectory (thick black line) and associated uncertainties (thin grey lines) for the streamer, as discussed in Section \ref{sec:modeling}, in 3D position--position--position space (RA, Decl., and LOS distance). 
    The blue disk and red cone represent the protoplanetary disk and its angular momentum orientation, respectively, as discussed in Section \ref{sec:disk}.
    An interactive version of the left panel is available online, allowing rotation and zooming of the 3D structures. 
    }
  \label{fig:morphology}
\end{figure}

Furthermore, the best-fit trajectory from TIPSY also allows us to reconstruct the three-dimensional geometry of the streamer, as shown in Figure \ref{fig:morphology}. Simulations \citep{Thies2011,Dullemond2019,Kuffmeier2021} and observations \citep{Ginski2021} have suggested that infalling streamers formed via cloudlet capture can be dynamically unrelated to the YSO, and thus induce misalignments in protoplanetary disks.
One way to investigate this is by comparing the orientation of the streamer with that of the disk, and as shown in Figure \ref{fig:morphology}, they appear moderately misaligned.
Quantitatively, the specific angular momentum vector ($\overrightarrow{r_{0}}\times\overrightarrow{v_{0}}$) will be perpendicular to the plane of the infalling gas and can be compared to the corresponding rotational axis of the disk, as discussed further in Section \ref{sec:disk}. 


\subsection{Streamer mass} \label{sec:mass}

Assuming that the C$^{18}$O (2--1) emission from structure C is optically thin, a reasonable assumption as flux ratio for C$^{18}$O to $^{13}$CO emission is roughly consistent with their abundance ratios \citep[$\sim7$,][]{Wilson1994}, the mass of the streamer ($M_{\text{streamer}}$) can be estimated as:
\begin{equation} \label{equ:mass}
M_{\text{streamer}}=\frac{2.37m_{H}4\pi D^{2}F_{\text{streamer}}}{A_{\text{C18O}}h\nu x_{\text{C18O}}f_{u}}    
,\end{equation}
\noindent where $m_{H}$ is the mass of a hydrogen atom, $A_{\text{C18O}}$ is the Einstein A coefficient of observed line transition, $\nu$ is the line frequency, $x_{\text{C18O}}$ is the abundance of the molecule relative to H$_{2}$, and $f_{u}$ is the fraction of molecules in the upper energy state of the transition \citep[e.g.,][]{Bergin2013}. 
Here, $f_{u}$ can be further computed as $f_{u} = 3e^{E_{u}/T}/Q_{\text{C18O}}(T)$, where $E_{u}$ is the upper state energy for the transition, $T$ is the gas temperature, and $Q_{\text{C18O}}(T)$ is the partition function for the molecule. We took $A_{\text{C18O}}$ and $E_{u}$ values to be 6.011 $\times 10^{-7}$~s$^{-1}$ and 15.81~K, respectively, from Leiden Atomic and Molecular Database \citep{lamda}. $x_{\text{C18O}}$ was taken as 1.8$\times 10^{-7}$ from \cite{Wilson1994}. We assumed the gas temperature ($T$) of $25\pm5$~K, typical at these scales \citep[e.g.,][]{Jorgensen2005}. This resulted in $Q_{\text{C18O}}(T)$ of $9.3\pm1.7$, estimated by interpolating between values provided in Cologne Database for Molecular Spectroscopy \citep{cdms}.

\begin{figure}[tbhp]
  \centering
  \includegraphics[width=0.45\textwidth]{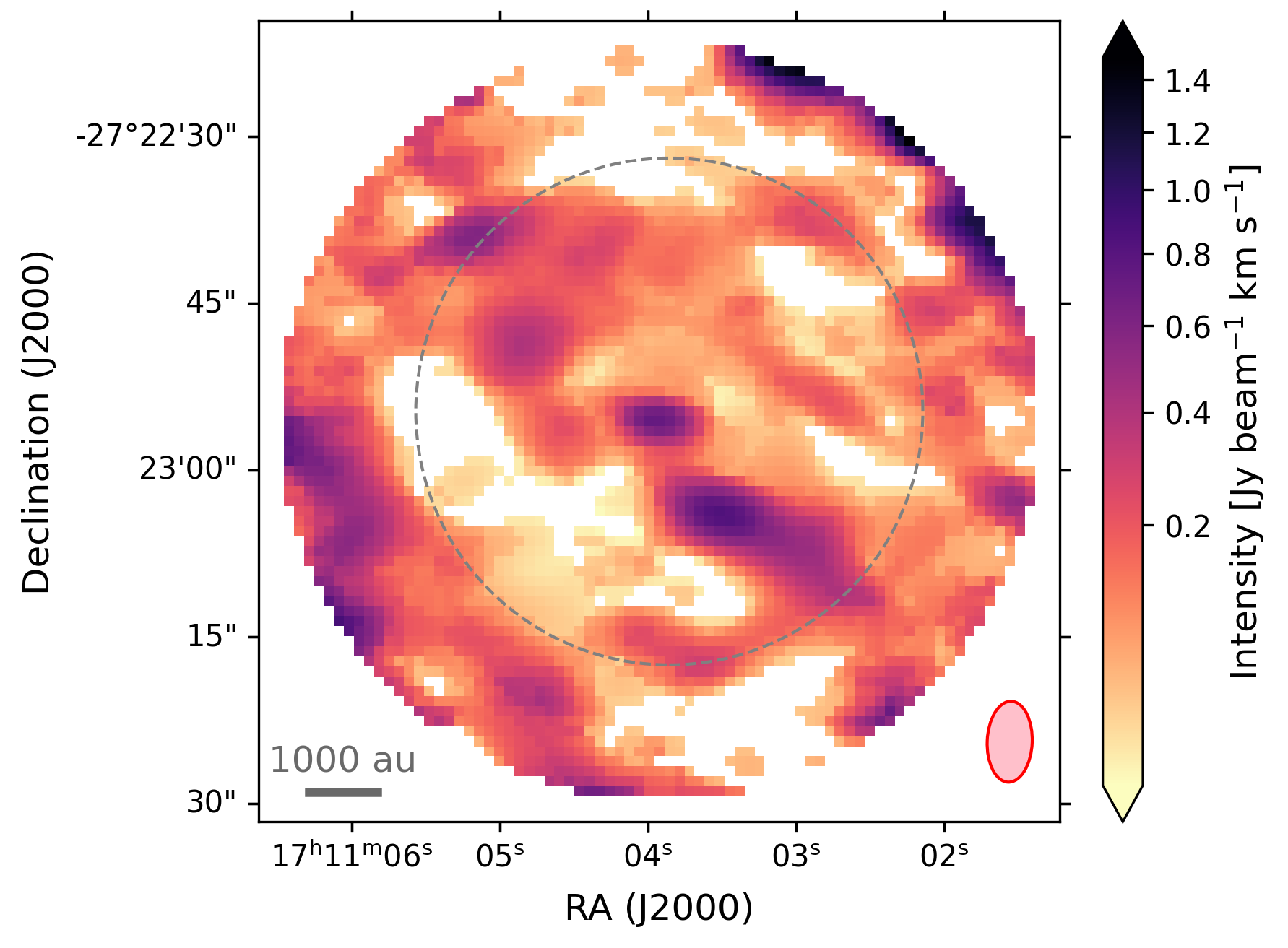}
    \caption{ACA (7m array) integrated intensity (moment 0) maps of C$^{18}$O (2--1) emission line.
    The horizontal grey line in the bottom-left corner represents a length scale of 1000 au, and the pink ellipse in the bottom-right corner represents the beam size.
    Grey dashed circle denotes the half power beam width of the primary beam.
    }
  \label{fig:aca_m0}
\end{figure}

For the primary mass estimation, we rely solely on the ACA data, as described in Section \ref{sec:obs} and shown in Figure \ref{fig:aca_m0}. This is because standalone ACA observations are the most sensitive to the length-scale ($\gtrsim20\arcsec$) of the streamer and have roughly twice the field-of-view. If we focus on the emission from the streamer, as seen in south-west quadrant of Figure \ref{fig:aca_m0}, we get a total flux of $\sim2.34\times10^{-20}$~W~m$^{-2}$. Following Equation \ref{equ:mass}, this corresponds to the total mass of streamer to be $(1.6\pm0.5)\times10^{-3}$~M$_{\odot}$ or $1.7\pm0.5$~M$_{\mathrm{Jup}}$. Multiplying the mass estimate with dynamical properties estimated using TIPSY, we estimated the total angular momentum and energy of the streamer as tabulated in Table \ref{tab:properties}. We note the ACA observations are still limited in field-of-view and largest recoverable scale, and thus, we may be missing substantial streamer emission. 
Moreover, we are not accounting for the effects of CO depletion.
Therefore, the mass and mass-derived properties should be taken as lower limits.

We cannot use the same streamer mass to estimate mass infall rate if our infall time is estimated for the smaller-scale C1+ACA data. This is because, unlike specific angular momentum and specific total energy, infall time depends on the exact point of the trajectory. From the TIPSY analysis of C1+ACA data presented in Figure \ref{fig:m0}, infalling timescale was estimated to be $17242\pm2003$~yr for gas at a distance of $\sim20\arcsec$ (roughly the field-of-view of C1+ACA data) from the protostar. The total flux from streamer, within these smaller field-of-view, is $\sim1.8\times10^{-20}$~W~m$^{-2}$, which corresponds to a mass of $1.3\pm0.4$~M$_{\mathrm{Jup}}$. 
This is in agreement with mass estimate for ACA standalone data (Figure \ref{fig:aca_m0}) within the same distance of $\sim20\arcsec$.
The average mass infall rate ($M_{\text{streamer}}/T_{\mathrm{inf}}$) for the streamer is then estimated as $(7.2\pm2.2)\times10^{-8}$~M$_{\odot}$~yr$^{-1}$ or $75\pm23$~M$_{\mathrm{Jup}}$~Myr$^{-1}$.

If we account for all the extended C$^{18}$O emission observed within the field-of-view of C1+ACA data, within the velocity range of 2--5~km~s$^{-1}$ where we expect most of the bound emission to lie (see Figure \ref{fig:channels}), we can similarly estimate the total mass of gaseous ``envelope" around BHB1. We found this mass to be $1.7\pm0.5$~M$_{\mathrm{Jup}}$, corresponding to the total flux of $\sim2.4\times10^{-20}$~W~m$^{-2}$. This suggests that the streamer contains $\sim80\%$ of total extended gas bound to BHB1. Moreover, since it is unclear whether the remaining extended emission is truly infalling or simply foreground or background cloud material, the actual fraction of mass within the streamer may be even higher. We note that this fraction reduces to $\sim30\%$ when considering the wider field-of-view of standalone ACA observations. However, that is primarily due to the contribution of bright emission at the edges of ACA field, which does not seem to be connected to BHB1 source (e.g., Figure \ref{fig:aca_m0}) and may instead represent interferometric artifacts, primary beam correction effects, or substructures in larger-scale cloud. 


\begin{deluxetable}{lc}[h!]
\tablecaption{Streamer Properties\label{tab:properties}}
\tablehead{
\colhead{Quantity} & \colhead{Value}
}
\startdata
Initial R.A. offset [au] & $-2414 \pm 235$ \\
Initial Dec. offset [au] & $-2225 \pm 267$ \\
Initial L.O.S. offset$^*$ [au] & $880 \pm 241$ \\
Initial R.A. speed$^*$ [km s$^{-1}$] & $0.35 \pm 0.16$ \\
Initial Dec. speed$^*$ [km s$^{-1}$] & $0 \pm 0$ \\
Initial L.O.S. speed [km s$^{-1}$] & $-0.21 \pm 0.17$ \\
Infall time [yr] & $17242 \pm 2003$ \\
Misalignment [deg] & $69 \pm 34$ \\
Mass$^{\dagger}$ [M$_{\mathrm{Jup}}$] & $1.66 \pm 0.51$ \\
Mass infall rate [M$_{\mathrm{Jup}}$/Myr] & $75 \pm 23$ \\
Angular Momentum$^{\dagger}$ [g cm$^2$ s$^{-1}$] & $(3.41 \pm 2.31) \times 10^{51}$ \\
Total energy$^{\dagger}$ [erg] & $(-1.56 \pm 0.54) \times 10^{40}$ \\
\enddata
\tablecomments{
$^*$ These quantities are fitted directly within TIPSY, as discussed in Section \ref{sec:modeling}.
$^{\dagger}$ These values depend on the estimated total mass of the streamer. As discussed in Section \ref{sec:mass}, we are likely underestimating the total mass and therefore these quantities should be considered lower limits.}
\end{deluxetable}

\subsection{Disk properties} \label{sec:disk}

To understand the role of streamers in enhancing mass-budget to form planetary systems, we aim to compare streamer mass and mass infall rates to disk mass. Our observations, as shown in Figure \ref{fig:m0}, provide us the total disk luminosity in all the three CO isotopologues. Thermochemical disk models have shown that disk masses can be constrained using multiple CO isotopologue line emission, particularly C$^{18}$O (2--1) and $^{13}$CO (2--1) \citep{Miotello2016}. The major drawback of these models is that they do not account for CO depletion and therefore, may underestimate the disk mass. However, this effect is expected to be minimal in disks around more massive Herbig stars ($\gtrsim1.5$~M${\odot}$), likely because they are warmer than typical T~Tauri disks \citep{Kama2020,Miotello2023}. As the stellar mass for BHB1 is $\sim2$~M${\odot}$ \citep{Alves2020}, it is expected to have a low CO depletion as well. 
Moreover, these limitations also persist in the mass estimation of the streamer (Section \ref{sec:mass}), and both sets of quantities should be taken as order of magnitude estimates. 
The total luminosity of the BHB1 disk is $\sim3.2\times10^5$~Jy~km~s$^{-1}$~pc$^{2}$ in C$^{18}$O (2--1) emission and $\sim1.2\times10^6$~Jy~km~s$^{-1}$~pc$^{2}$ in $^{13}$CO (2--1) emission. Using equation 2 in \cite{Miotello2016}, the corresponding disk mass ($M_{\mathrm{disk}}$) is $3.8$~M$_{\mathrm{Jup}}$. 

Using the mass of the disk, we can also try to estimate its angular momentum and compare it to the streamer's. To do this properly, we need to know the surface density profile of the disk and then multiply it by the Keplerian rotation. However, the resolution of observations is not high enough to characterize the surface density profile. 
As an approximation, we can assume that radius of gyration ($k$, effective radius for angular momentum estimation) is roughly half of the actual disk radius.
From visual inspection of the CO (2--1) observations \citep[Fig. 1 in][]{Alves2020}, the disk radius appears to be $\sim2\arcsec$, corresponding to $\sim320$~au at a distance of 163~pc. Using $k\approx160$~au, the angular momentum can then be estimated as $M_{\mathrm{disk}}\sqrt{GM_{*}k}$, yielding a value of $6\times10^{51}$~g~cm$^2$~s$^{-1}$. This is remarkably consistent with the streamer's estimate within uncertainties.
We emphasize that this is only an order-of-magnitude estimate and an accurate determination would require characterising the surface density, as done by \citet{Longarini2024}.



As angular momentum is a vector quantity, it is also worth comparing the orientation of the disk rotation with respect to the streamer. As discussed in Section \ref{sec:modeling}, we already know the orientation for the streamer via TIPSY. For the orientation of the disk, we adopt the values fit from the continuum visibilities presented in \cite{Alves2020}, which yield a position angle (PA) and inclination ($i$) of $14.4\pm0.5^\circ$ (measured east of north for the redshifted side) and $76\pm2^\circ$, respectively. Using these orientation angles, the 3D orientation of disk rotational axes can be constrained as
\begin{align*}
\hat{x}_{\mathrm{disk}} & = (1 - \cos^2(i)) \cos(\mathrm{PA})\,\hat{i} \quad\text{(along RA)} \\
\hat{y}_{\mathrm{disk}} & = -(1 - \cos^2(i)) \sin(\mathrm{PA})\,\hat{j} \quad\text{(along Dec.)}\\
\hat{z}_{\mathrm{disk}} & = \pm \cos(i)\,\hat{k} \quad\text{(along LOS)}
\end{align*}
Here, the ``$\pm$'' factor in the estimation of $\hat{z}_{\mathrm{disk}}$ reflects the degeneracy in whether the angular momentum vector is pointed towards or away from the observer. This degeneracy can be resolved by identifying the near side of the disk. Scattered light images of the disk reported in \cite{Zurlo2021} (Figure 2) show a clear bowl-like morphology of the disk's surface layer to the east, and the emission from the west side of the disk appears significantly fainter. This suggests that the eastern side of the disk is facing the observer \citep[see][for more resolved examples]{Avenhaus2018}. Since the velocity maps (Figure \ref{fig:m1}) indicate that the rotational axis is pointing toward the east, we infer that $\hat{z}_{\mathrm{disk}} = -\cos(i)$. Therefore, the orientation of disk's angular momentum can be represented as $0.91\pm0.02\,\hat{i} -0.23 \pm 0.01\,\hat{j} -0.24 \pm 0.03\,\hat{k}$. Comparing it to streamer's orientation (Section \ref{sec:modeling}), we found these two vectors to be misaligned by $69\pm34^\circ$, as shown in Figure \ref{fig:morphology}.


\section{Discussion} \label{sec:discussion}

\subsection{Infalling streamer} \label{sec:streamer}

 Contrary to the structures A and B, the structure C to the west of BHB1 appears to be bound to the system, as shown in Figure \ref{fig:channels}. Moreover, we were able to simultaneously fit both its morphology and velocity profile with infalling trajectories using TIPSY, as discussed in Section \ref{sec:modeling}. Although no assumptions were made regarding the energetics of these trajectories, they were found to lie on bound orbits with negative total energy. These results strongly suggest that this structure is indeed a streamer infalling onto the BHB1 system.

As discussed in Section \ref{sec:disk}, the streamer seems to be moderately misaligned ($\sim69^\circ$) with the disk. This suggests that the streamer is dynamically unrelated to the original protostellar system. Therefore, it may represent a ``cloudlet-capture" event, where the streamers are induced due to interaction with large-scale clouds \citep[e.g.,][]{Dullemond2019, Hanawa2024} and not due to substructures within the natal gaseous envelopes \citep[e.g.,][]{Tu2024}. This interpretation is further supported by the fact that $\sim80\%$ of the extended gas bound to the BHB1 system seems to be within this streamer, as discussed in Section \ref{sec:mass}. 
The dynamical state of the remaining extended gas is unclear, and it may not even be infalling onto the system. 
While \citet{Alves2020} found evidence of foreground absorption around the systemic velocity in CO (2--1), it appears negligible in C$^{18}$O (2--1). 
Moreover, our ACA observations have largest recoverable scale of $\sim30\arcsec$ ($\sim5000$~au), larger than the typical envelopes of $\sim1000$ au around Class I sources \citep[e.g.,][]{Heimsoth2022}, so it is unlikely that significant envelope emission has been filtered out.
This would imply that most, if not all, of the original natal envelope has already been dispersed, and what we observe now is a rejuvenated infall event triggered by interactions between BHB1 and its surrounding molecular cloud.
Although the source BHB1 has been classified as a flat-spectrum source, it could be a Class II source which looks slightly less-evolved due to a highly inclined disk \citep[e.g.,][]{Dunham2014} and late-stage infall via streamer \citep[e.g.,][]{Kuffmeier2023}.

Theoretical studies \citep[e.g.,][]{Thies2011,Dullemond2019,Kuffmeier2021,Kuffmeier2024,Pelkonen2025} have suggested that infall of dynamically unrelated material, as we think is happening to BHB1, can misalign the disks.
This idea has been further supported by detection of streamers around sources with misaligned disks \citep[e.g.,][]{Ginski2021,Tanious2024}. 
However, an observational study comparing angular momentum of disk and streamer falling onto it was missing. 
We show that the streamer is not only misaligned with the disk, but also carries a comparable amount of angular momentum (on the order of a few $10^{51}$~g~cm$^2$~s$^{-1}$), as inferred in Sections \ref{sec:mass} and \ref{sec:disk}.
This suggests that the interaction with the environment can be a natural way to induce disk misalignments, which can further explain misalignments observed in a significant fraction ($\gtrsim30\%$) of planetary systems \citep[e.g.,][]{Albrecht2022,Biddle2025}.

Another interesting property of this streamer is that its mass ($\sim1.7$~M$_{\mathrm{Jup}}$, Section \ref{sec:mass}) is comparable to the mass of protoplanetary disk ($\sim3.8$~M$_{\mathrm{Jup}}$, Section \ref{sec:disk}) around BHB1. This means that this streamer can significantly enhance the mass available to form planets. This further suggests that cloudlet capture events can be a viable resolution to the apparent ``mass-budget problem" of Class II disks, where the disks do not seem to have enough material to form the observed population of planetary systems \citep[e.g.,][]{Manara2018,Mulders2021}. As even our standalone ACA observations have limited recoverable scale and field-of-view, the streamer mass estimates should be taken as a lower limit on the total mass that can be supplied by the environment. 

We estimated the mean mass infall rate for the streamer to be 
$\sim75$~M$_{\mathrm{Jup}}$~Myr$^{-1}$, as discussed in Section \ref{sec:mass}. Comparison with the estimated disk mass shows that such an infall can double the mass-budget to form planets within $\sim0.05$~Myr, which is a small fraction of the typical disk lifetime of a few Myr \citep[e.g.,][]{Pfalzner2022}. This suggests that streamers need not be commonly observed to be significant, as even short-lived infall episodes can strongly impact disk evolution, leaving many systems affected but without present-day signatures.

\begin{figure}[tbhp]
  \centering
  \includegraphics[width=0.45\textwidth]{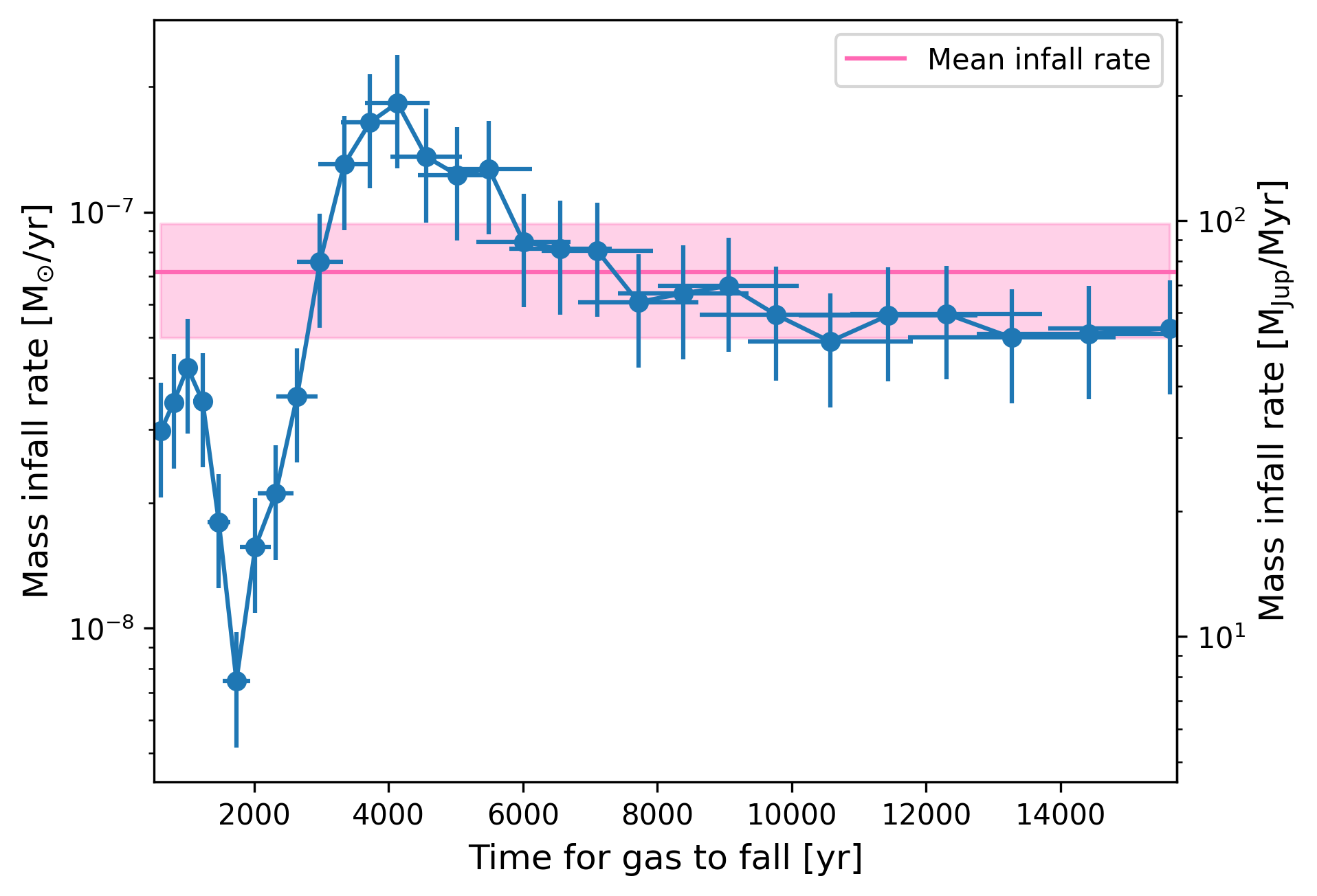} 
    \caption{Expected variation of mass infall rate (y-axis) as a function of time (x-axis) due to observed substructures within the streamer, as discussed in Section \ref{sec:mass}.
    The pink line and associated shaded region represents the mean infall rate for the streamer and the uncertainty on it. 
    }
  \label{fig:mass_var}
\end{figure}

Furthermore, as discussed in Section \ref{sec:mass} and shown in Figure \ref{fig:mass_var}, the mass infall rate from the streamer varies by up to one order of magnitude. Such variability can have important implications for triggering disk instabilities, like thermal or gravitational instabilities, which can further induce mass accretion outbursts onto protostars and even trigger planetesimal formation \citep[e.g.,][]{Clarke1996,Lodato2004,Dunham2012,Padoan2014,Bae2014,Jensen2018,Calcino2025,Longarini2025}. Interestingly, an infalling streamer has been recently reported around FU Ori outbursting system \citep{Hales2024}. Variability in mass infall rate also implies that the absence of observable infall signatures in some systems does not necessarily indicate a lack of infall altogether; they may simply be observed during a quiescent phase, when the infall rate is briefly low.
%

\subsection{Origin of structures A and B} \label{sec:mys}

As shown in Figure \ref{fig:channels} and discussed in Section \ref{sec:analysis}, the red-shifted structure to the north (labelled A) and blue-shifted structure to the south (labelled B) seem unbound from the BHB1 system, suggesting that they are not the typical infalling streamers that are bound to the protostar \citep[e.g.,][]{Thieme2022,Pineda2023}. Despite that, the fact that they seem connected to the disk and even aligned with disk velocities on their respective sides suggests that they are not chance alignment of substructures within large-scale cloud with the BHB1 system. Here we speculate on some of the possible mechanisms causing them:
\begin{itemize}

\item \textit{Outflows}: As these structures exhibit opposite velocities, they resemble bipolar outflows seen commonly around young sources \citep[e.g.][]{Tychoniec2019}. However, ejection of material seen in other observations as well as simulations \citep[e.g.][]{Vorobyov2020} generally continues away from the central source and does not curve back, as these structures appear to do. It may be because these outflows are being gravitationally captured by the [BHB2007]-2, hereafter BHB2, system nearby, as detected in near-infrared images presented in \citet{Zurlo2021}. However, \citet{Zurlo2021} estimated the mass of BHB2 system to be $\sim1$~M$_{\odot}$, roughly a factor of two lower than the mass of BHB1. Therefore, it is unclear how this high velocity material can be trapped by a less massive system. Alternatively, it could be that the shape of the outflow is more directly affected by the feedback from the BHB2 \citep[e.g.,][]{Yen2019}, or due to interaction with surrounding larger-scale clouds \citep{Duarte-Cabral2012}. 

As outflows are generally expected to be parallel to the rotational axes of YSOs \citep[e.g.,][]{Gupta2022}, we can also examine the orientation of these structures relative to the disk of BHB1. 
As discussed in Section \ref{sec:disk}, the disk inclination and position angle are estimated to be $76^{\circ}$ and $14.4^{\circ}$, respectively. 
This implies that the disk is almost edge-on with its rotational axis roughly aligned to the East-South direction, contrary to the orientation of structures A and B. It is also possible that the outflows originate from an inner disk that is misaligned with the outer disk observed by ALMA. However, \citet{Zurlo2021} reports a tentative jet-like emission visible in NIR wavelengths consistent with the rotational axis of the outer disk, and thus perpendicular to the structures A and B.

\item \textit{Interactions with nearby sources}: The observed tail-like structures may represent interactions with nearby systems BHB2 and VLT 171104.07-272258.8. They both are located a few arcseconds south-east of the BHB1 \citep{Zurlo2021}, between the BHB1 source and region where structures A and B converge again. According to the proper motion analysis using NIR observations, \citet{Zurlo2021} demonstrated that VLT 171104.07-272258.8 is not comoving with BHB1, and is likely to be a background object. On the other hand, BHB2 is more likely to be physically associated with BHB1 and may be interacting with it. 

A typical interaction in the form of a stellar flyby can induce double-sided tidal tails \citep[e.g.,][]{Cuello2023}. 
However, these tails are usually curved in the opposite direction, with one tail typically extending towards the perturbing source while the other pointing away. This is inconsistent with the observed morphology of structures A and B, both of which curve eastward towards BHB2.
The dynamical signatures of these structures also differ from those expected in both prograde and retrograde encounters, as shown in \cite{Cuello2020}, although these signatures can be sensitive to the system’s inclination relative to the observer.

\item \textit{Feedback from BHB2}: Structures A and B may simply be overdense regions within the larger hourglass-shaped gas distribution located east of BHB1. The region within this hourglass-like morphology is generally depleted in gas and contains the BHB2 source. It is possible that feedback from BHB2, likely in the form of mass ejection since the source appears too low-mass to drive significant radiative feedback, has cleared out the gas and contributed to the formation of these structures. 
Interestingly, we do not detect any disk emission from BHB2 itself, which may further suggest that a violent ejection dispersed the disk.
However, in this scenario, it remains unclear why structures A and B, if shaped solely by feedback from BHB2, are so well aligned both spatially and kinematically with BHB1.

\item \textit{Hyperbolic infall}: As discussed above, these structures appear to be unbound from the system. Although almost all of the known streamers appear to be in bound parabolic or elliptical orbits \citep[e.g.,][]{Pineda2023}, in theory, they can still be falling towards the source on unbound hyperbolic orbits. This scenario has to have a very specific orientation where the initial velocity of infall gas is pointed directly at the protostellar system. In such a scenario, infalling gas may then hit the disk and lose kinetic energy due to shocks \citep[e.g.,][]{Garufi2024}. However, as pointed out in \citet{Alves2020}, these structures exhibit very little velocity gradient, which is unlikely for a hyperbolic unless the plane of motion is roughly aligned with the POS. 
To test this further, morphology and velocity gradient of the streamer could be fit simultaneously using a streamer fitting code like TIPSY 
\citep{Gupta2024}. However, CO (2--1) emission is largely absorbed or filtered out at velocities close to systemic velocity ($\sim2.5$--4.5~km~s$^{-1}$) of the protostellar source, which are the key to such fitting procedures. Higher sensitivity observations are required to detect these structures in more optically thin isotopologues and test this hypothesis. 

\end{itemize}

\section{Summary}  \label{sec:sum}

The recent ALMA observations of the BHB1 system reveal two puzzling structures (A and B) that exhibit red- and blue-shifted velocities and appear spatially and kinematically connected to the planet-forming disk. However, as they seem to be unbound from the system, they are not typical infalling streamers and their origin remains ambiguous. Several possibilities are considered, including outflows, interactions with nearby BHB2 system, and hyperbolic infall. However, none of these scenarios fully account for all the observed properties of these structures, suggesting that multiple dynamical processes may be at play. Thus, the origin of these two structures remain unknown.

In contrast, a third structure (C) on the western side of BHB1 is well explained by a bound infalling trajectory, consistent with typical infalling streamers. Its misalignment with the disk’s rotational axis and lack of a prominent gaseous envelope around BHB1 suggest that this streamer was induced by gravitational capture of material from surrounding clouds, and maybe represents a ``late-stage infall" of material onto a Class II disk. Interestingly, both the mass and angular momentum of the streamer are comparable to the planet-forming disk, suggesting that such events can significantly enhance their mass budget to form planets and reshape the disk orientations. Overall, these results show that star-formation environments continue to significantly influence star and planet formation processes, long after dispersal of natal dense cores. 


\section*{Acknowledgments}

LIC and AG acknowledge support from NSF AST-2407547 and the David and Lucile Packard Foundation and the Virginia Institute of Theoretical Astronomy (VITA).
A.Z. acknowledges support from ANID -- Millennium Science Initiative Program -- Center Code NCN2024\_001 and Fondecyt Regular grant number 1250249.
This project has received funding from the European Research Council (ERC) under the European Union Horizon Europe programme (grant agreement No. 101042275, project Stellar-MADE).
ZYL is supported in part by NASA 80NSSC20K0533, NSF AST-2307199, and the Virginia Institute of Theoretical Astronomy (VITA).
T.B. acknowledges financial support from the FONDECYT postdoctorado project number 3230470.
JMG acknowledge support by the grant PID2023-146675NB-I00 (MCI-AEI-FEDER, UE) and  by the program Unidad de Excelencia María de Maeztu CEX2020- 001058-M. 

This paper makes use of the following ALMA data: ADS/JAO.ALMA\#2023.1.01065.S. ALMA is a partnership of ESO (representing its member states), NSF (USA) and NINS (Japan), together with NRC (Canada), MOST and ASIAA (Taiwan), and KASI (Republic of Korea), in cooperation with the Republic of Chile. The Joint ALMA Observatory is operated by ESO, AUI/NRAO and NAOJ.


%
\facilities{ALMA}

\software{astropy \citep{astropy:2013, astropy:2018, astropy:2022},  
TIPSY \citep{Gupta2024}
          }


\appendix

\section{Fitting uncertainties} \label{sec:erros}

\begin{figure*}[htbp]
  \centering
  \includegraphics[width=0.95\textwidth]{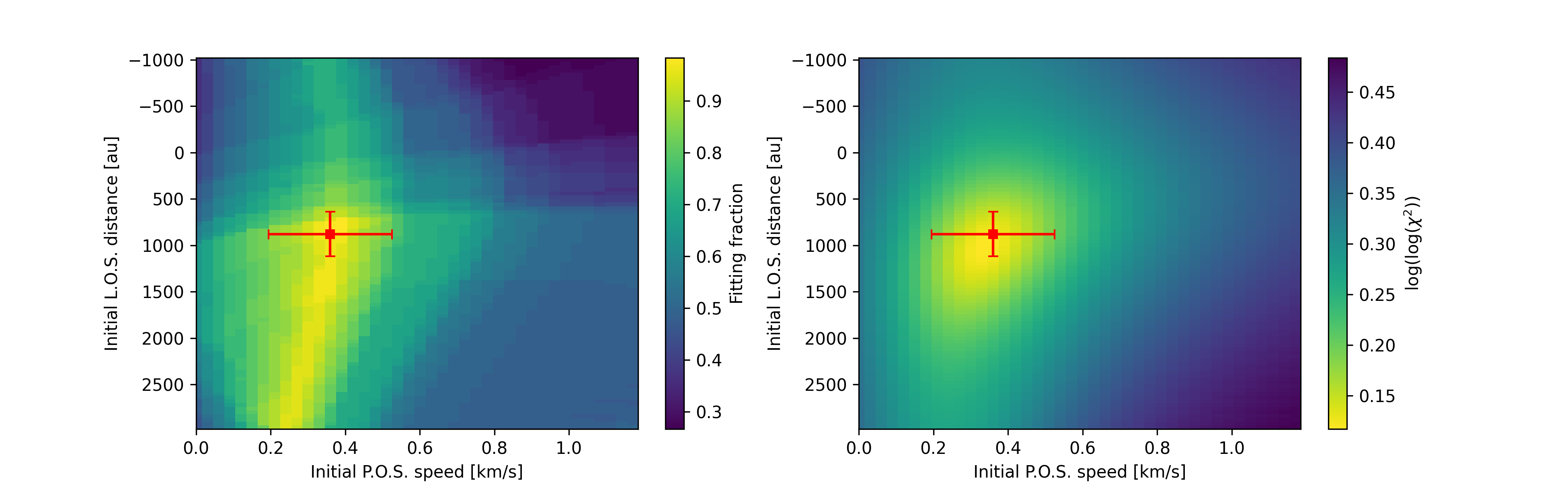}
    \caption{Distribution of goodness-of-fit estimates as functions of free parameters: initial speed on the POS (x-axis) and initial spatial offset in the LOS direction, for TIPSY fitting for structure C. Here, initial direction of gas in the POS (third free parameter) is fixed to the value for the best fit.
    \textit{Left panel}: Distribution of fractions of coordinate values of points in the observed streamer curve (intensity-weighted means and standard deviations), which is consistent with the theoretical trajectories. 
    \textit{Right panel}: Distribution of $\log(\log(\chi^{2}))$ deviation between the observed streamer curve and theoretical trajectories. 
    In both plots, yellow regions represent good fits.
    }
      \label{fig:errors}
\end{figure*}

For the dynamical modeling of structure C using TIPSY, the error estimation procedure differed slightly from the default approach described in \citet{Gupta2024}. By default, errors are estimated using trajectories fitting $>90\%$ of the intensity-weighted mean coordinates of observed streamer emission \citep[for more details, see Section 2.3,][]{Gupta2024}. However, we found this method to overestimate the uncertainties when comparing the resulting trajectories to the observed morphology and velocity gradient of the streamer emission, similar to Figure~\ref{fig:tipsy}.
Instead, the uncertainties reported in Table~\ref{tab:properties} and shown in Figure~\ref{fig:errors} were derived from fits with a $\chi^2$ deviation $<27$. For comparison, the best-fit model has a $\chi^2$ value of 21. The error bars (in red, Figure~\ref{fig:errors}) represent the standard deviation of the parameters (240~au for initial L.O.S. distance, 0.16~km~s$^{-1}$ for initial P.O.S. speed) for these reasonable fits.


\bibliography{refs}{}
\bibliographystyle{aasjournalv7}



\end{document}